\documentclass{svjour3}                     
\smartqed  
\usepackage{graphicx}
\begin{document}

\title{Orbital evolution of P\v{r}\'{i}bram and Neuschwanstein
}

\author{Leonard Korno\v{s} \and Juraj T\'{o}th \and Peter Vere\v{s} 
}
\institute{\at
              Department of Astronomy, Physics of the Earth and Meteorology\\
              Faculty of mathematics, physics and informatics\\
              Comenius University\\
              Mlynsk\'{a} dolina\\
              842 48 Bratislava\\
              Slovak Republic
               \\
              Tel.: +421-02-60295610\\
              \email{toth@fmph.uniba.sk}\\
}
\date{Received: date}

\maketitle
\begin{abstract}
The orbital evolution of the
two meteorites P\v{r}\'{i}bram and Neuschwanstein on
almost identical orbits and also several thousand clones were studied in
the framework of the N-body problem for 5000 years into the past. The
meteorites moved on very
similar orbits during the whole investigated interval. We have also searched
for photographic meteors and asteroids moving on similar orbits. There were
5 meteors found in the IAU MDC database and 6 NEAs with currently similar
orbits to P\v{r}\'{i}bram and Neuschwanstein. However, only one meteor 161E1
and one asteroid 2002 QG46 had a similar orbital evolution over the last
2000 years.


\end{abstract}

\section{Introduction}

\label{intro} It is almost 50 years since the fall (April 7, 1959) and recovery
of the P\v{r}\'{i}bram meteorite (Ceplecha 1961), the first meteorite with a
precisely known heliocentric orbit (Tab. 1). Later, the fall of the
Neuschwanstein meteorite
was observed on April 6, 2002 and it was successfully recovered (Oberst
et al. 2004). It was shown that both meteorites were moving on similar orbits
(Spurn\'y et al. 2003), but the question about their origin remains unanswered.
Moreover, their different meteoritic types, P\v{r}\'{i}bram being an
H5 ordinary
chondrite (Ceplecha 1961) with cosmic-ray exposure age 12 Myr (Stauffer and
Urey 1962) and Neuschwanstein an EL6 enstatite chondrite with cosmic-ray
exposure age 48 Myr (Bishoff and Zipfel 2003; Zipfel et al. 2003), makes their
common origin very problema\-tic. It is a challenge for the
scientific community to
explain the dynamical and physical evolution of these two meteorites. Earlier,
the existence of asteroidal-meteoritic streams was suggested by Halliday et al.
(1990). Recently, the observation of Neuschwanstein led Spurn\'{y} et al.
(2003) to suggest a heterogeneous meteoritic stream in the orbit of
P\v{r}\'{i}bram. On the other hand, a statistical analysis by Pauls and Gladman
(2005) showed that the occurrence of pairs as close as P\v{r}\'{i}bram and
Neuschwanstein is at the $10\%$ level, which is consistent with random chance.
Recently, Jones and Williams (2007) studied the
possible existence of meteoritic
streams. Trigo-Rodr\'{i}guez et al. (2007), performing orbital and spectral
analyses, found three meteorite-dropping bolides, which may well be associated
with the Near Earth
Asteroid 2002 NY40. In the present paper, we analyze possible
associations of meteors and NEAs with P\v{r}\'{i}bram and Neuschwanstein and
also their orbital evolutions on a time scale of 5000 years. Also, we discuss
the possible common origin of P\v{r}\'{i}bram and Neuschwanstein.

\begin{table}[t]
 \caption{Orbital elements (eq. 2000.0) of P\v{r}\'{i}bram and Neuschwanstein (Spurn\'y et al. 2003).}
 \centering
 \label{tab:1}
 \begin{tabular}{lcc}
 \hline\noalign{\smallskip}
   & P\v{r}\'{i}bram &  Neuschwanstein \\
 \tableheadseprule\noalign{\smallskip}
     ~$a$ & \hspace{22pt} 2.401 $\pm$ 0.002 AU  & \hspace{23pt} 2.40 $\pm$ 0.02 AU\\
     ~$e$ & \hspace{5pt} 0.6711 $\pm$ 0.0003    & \hspace{6pt} 0.670 $\pm$ 0.002\\
     ~$q$ & \hspace{22pt} 0.78951 $\pm$ 0.00006 AU & \hspace{23pt} 0.7929 $\pm$ 0.0004 AU\\
     ~$Q$ & \hspace{22pt} 4.012 $\pm$ 0.005 AU   & \hspace{24pt} 4.01 $\pm$ 0.03 AU \\
     ~$\omega$ & \hspace{0pt} 241.750$^\circ \pm$ 0.013$^\circ$   & \hspace{0pt} 241.20$^\circ\pm$ 0.06$^\circ$    \\
     ~$\Omega$ & \hspace{4pt} 17.79147$^\circ\pm$ 0.00001$^\circ$ & \hspace{5pt} 16.82664$^\circ\pm$ 0.00001$^\circ$\\
     ~$i$ & \hspace{0pt} 10.482$^\circ~\pm$ 0.004$^\circ$           & \hspace{5pt} 11.41$^\circ\pm$ 0.03$^\circ$\\
 \noalign{\smallskip}\hline
 \end{tabular}
 \end{table}

\section{Associations with P\v{r}\'{i}bram and Neuschwanstein}

The heliocentric orbits of P\v{r}\'{i}bram and Neuschwanstein are
almost identical (Tab. 1), but the errors in the orbital elements of
Neuschwanstein are about 1 order of magnitude larger compared to
P\v{r}\'{i}bram. However, both orbits are relatively precise and
the D-criterion of Southworth and Hawkins (1963), $D_{SH}=0.03$,
indicates a very close similarity.

We have searched for possible members of a meteoroid stream, to be
associated with the meteorites, in the IAU Meteor Database of
Photographic Orbits (Lindblad et al. 2003) based on $D_{SH}\leq
0.2$ (cf. Jones et al. 2006). There were 5 meteoroids found, which
are listed in Table 2 (for details of the designations see Neslu\v{s}an,
2003) and compared to the orbit of P\v{r}\'{i}bram. While the
P\v{r}\'{i}bram and Neuschwanstein entry masses were several hundred
kilograms, the other meteoroids mentioned in Table 2 are very small.
The photometric mass of the largest one, 161E1, is about 2100 g.

\tabcolsep=3.5pt
\begin{table}[t]
 \caption{Orbital elements (eq. 2000.0), geocentric velocity $V_{g}$, geocentric radiant
 ($RA$ and $DC$), magnitude and D-criterion of P\v{r}\'{i}bram and
 Neuschwanstein meteorites (Spurn\'y et al. 2003) as well as 5 meteoroids
 from the IAU Meteor Database
 (Lindblad et al. 2003).}
 \centering
 \label{tab:2}
 \begin{tabular}{lcccccccccccc}
 \hline\noalign{\smallskip}
meteo- &$q$&$a$&$e$&$i$&$\omega$&$\Omega$&$\pi$&$V_{g}$&$RA$&$DC$&mag&$D_{SH}$       \\
roid   &(AU)&(AU)&&($\circ)$&($\circ)$&($\circ)$&($\circ)$&(km/s)&($\circ)$&($\circ)$& &     \\

 \tableheadseprule\noalign{\smallskip}
 P\v{r}\'{i}br &0.790&2.401& 0.671& 10.5& 241.8& 17.8& 259.5& 17.43& 192.3& 17.5& -19.2 & -\\
 Neusch &0.793& 2.400& 0.670& 11.4& 241.2& 16.8& 258.0& 17.51& 192.3& 19.5& -17.2& 0.03 \\
 012F1&  0.776& 2.217& 0.650&  0.7& 244.6& 16.6& 261.3& 16.41& 183.3&  0.2&  -6.7& 0.17 \\
 161E1&  0.817& 2.696& 0.697&  9.6& 236.5& 18.9& 255.4& 16.95& 189.5& 17.8& -10.8& 0.06 \\
 079H1&  0.863& 2.757& 0.687&  8.9& 228.7& 19.8& 248.4& 15.43& 185.4& 20.6&   2.4& 0.15 \\
 130F1&  0.774& 2.867& 0.730& 16.1& 242.5& 20.2& 262.7& 19.93& 200.3& 22.6& -10.7& 0.12 \\
 083H1&  0.821& 2.582& 0.682&  4.9& 236.5& 21.7& 258.1& 16.01& 186.9&  8.6&   2.0& 0.10 \\
\noalign{\smallskip}\hline
\end{tabular}
\end{table}

Also we have searched for a possible parent body among Near Earth Asteroids. We
have found 6 NEAs from the current (April 2007) Bowell (2007) database, within
$D_{SH}\leq 0.2$. The osculating orbital elements compared to P\v{r}\'{i}bram
are listed in Table 3.

Similarity of osculating orbits is not enough to prove any
association among the orbits mentioned above. Therefore we have looked
for similarity in orbital evolution over the past 5000 years.
We have numerically integrated the motion of the P\v{r}\'{i}bram and
Neuschwanstein meteorites, the 5 meteoroids and 6 NEAs using the
multi-step procedure of the Adams-Bashforth-Moulton 12th order method,
with a variable step-length. The positions of the perturbing major
planets were obtained from the JPL Ephemeris DE406.

\tabcolsep=3.5pt
\begin{table}[t]
\caption{Orbital elements (eq. 2000.0) of  P\v{r}\'{i}bram
(Spurn\'y et al. 2003) as well as 6 objects from the NEA database
(Bowell 2007). H(1,0) is the absolute magnitude of NEAs and
$D_{SH}$ is the D-criterion.} \centering
 \label{tab:3}       
\begin{tabular}{lccccccccc}
 \hline\noalign{\smallskip}
name &$q$&$a$&$e$&$i$&$\omega$&$\Omega$&$\pi$&$H(1,0)$&$D_{SH}$    \\
\tableheadseprule\noalign{\smallskip} P\v{r}\'{i}bram & 0.790& 2.401& 0.671&
10.5& 241.8&  17.8& 259.5&     &       \\ 1998 SJ70 & 0.656& 2.236& 0.706& 7.4&
244.4&  23.8& 268.2& 18.3& 0.18  \\ 2002 EU11 & 0.746& 2.397& 0.689&  2.9&
274.5& 346.3& 260.8& 20.9& 0.15
\\ 2002 QG46 & 0.905& 2.434& 0.628&  8.3& 268.2& 346.0& 254.2& 19.6& 0.17  \\ 2003 RM10 & 0.755& 1.847& 0.591&
13.7& 287.0& 341.6& 268.6& 20.2& 0.20  \\ 2005 GK141& 0.938& 2.735& 0.657& 14.0& 218.2&  34.2& 252.5& 22.1& 0.19
\\ 2005 RW3  & 0.754& 2.107& 0.642&  2.7& 218.9&  49.4& 268.3& 22.8& 0.18  \\
 \noalign{\smallskip}\hline
 \end{tabular}
 \end{table}

\begin{figure*}
\centering
  \includegraphics[width=0.33\textwidth, angle=-90]{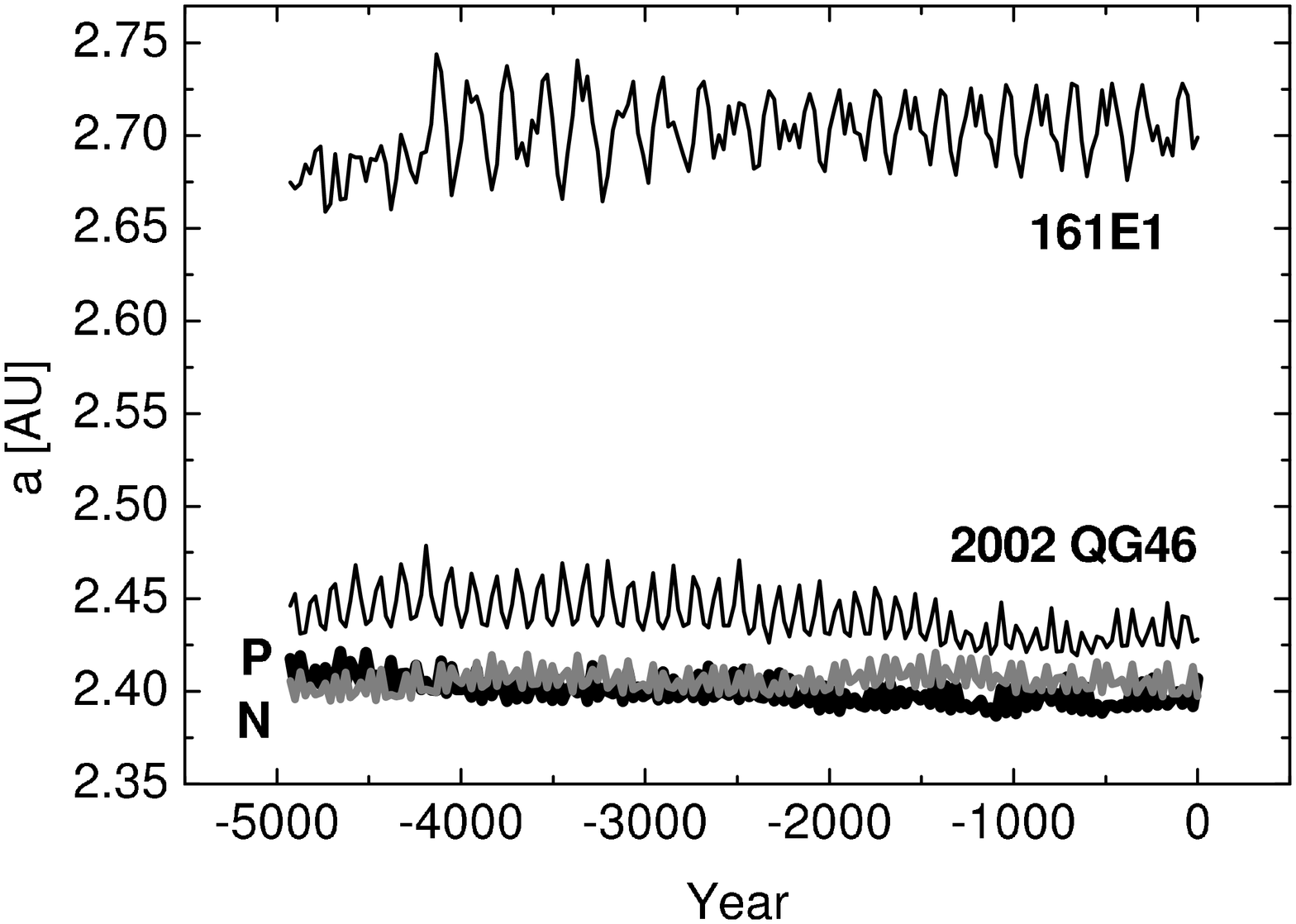}
\vspace{0.3cm} \centering \hspace{0.2cm}
  \includegraphics[width=0.33\textwidth, angle=-90]{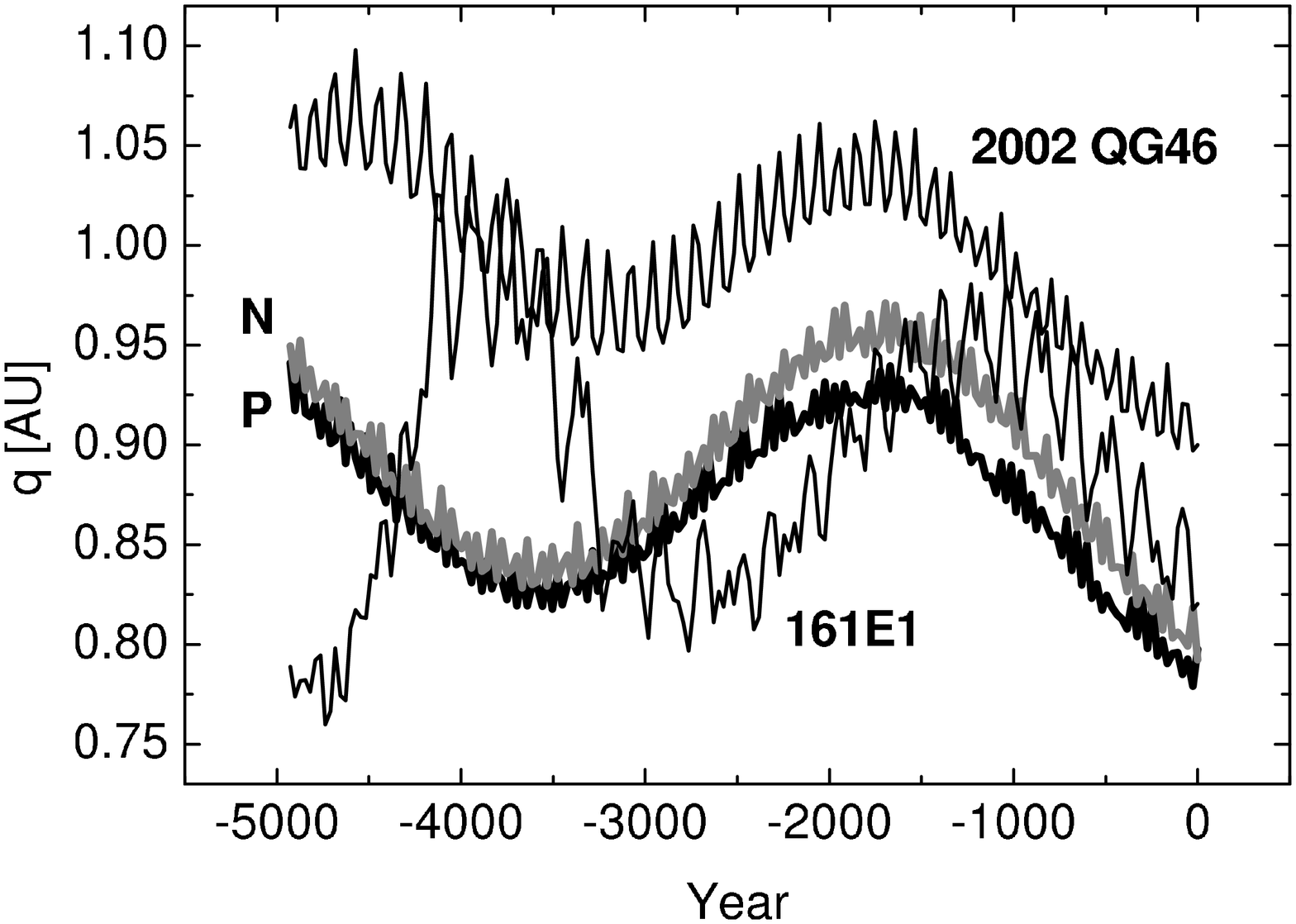}
  \includegraphics[width=0.33\textwidth, angle=-90]{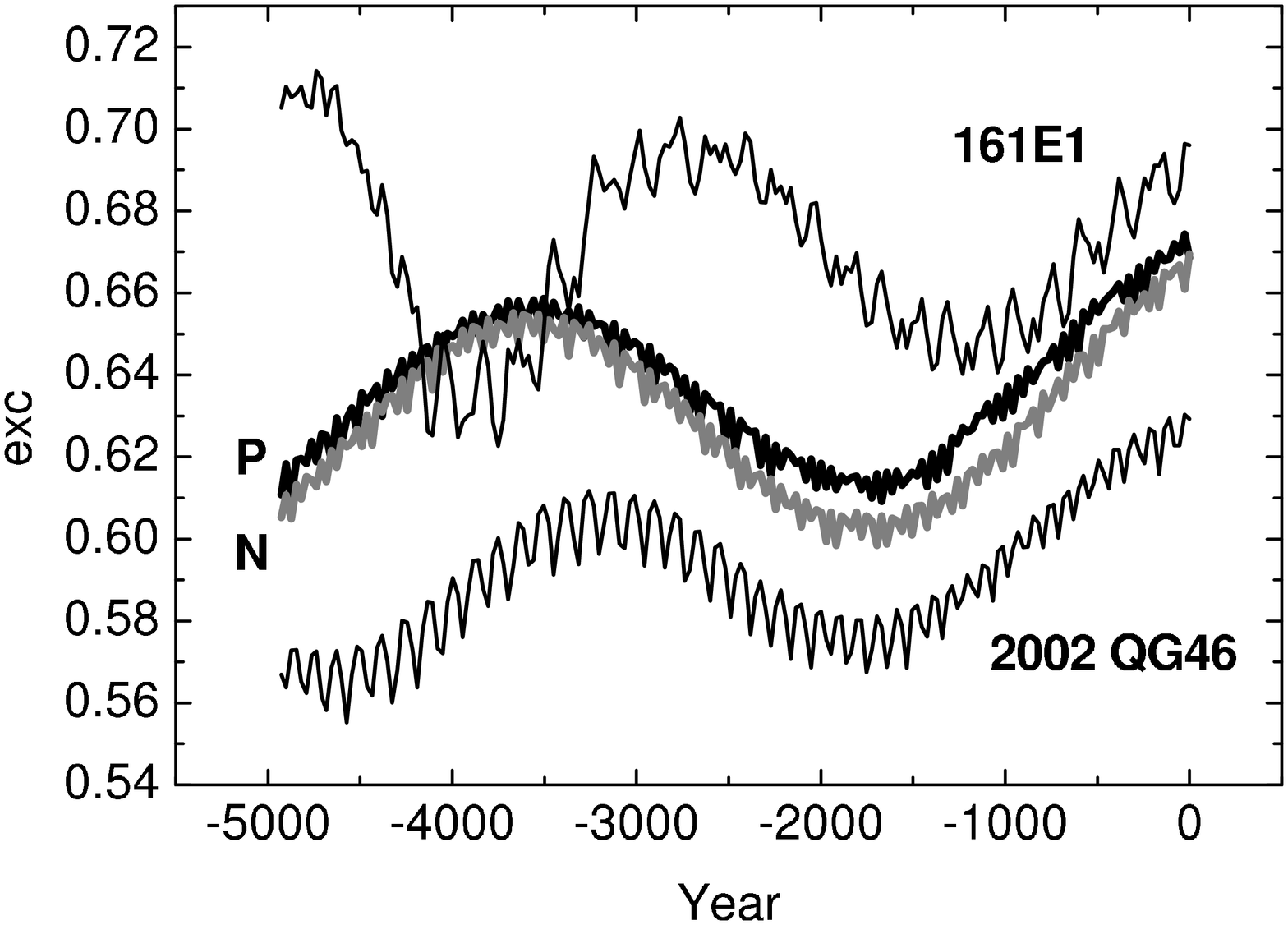}
\vspace{0.3cm} \centering \hspace{0.2cm}
  \includegraphics[width=0.33\textwidth, angle=-90]{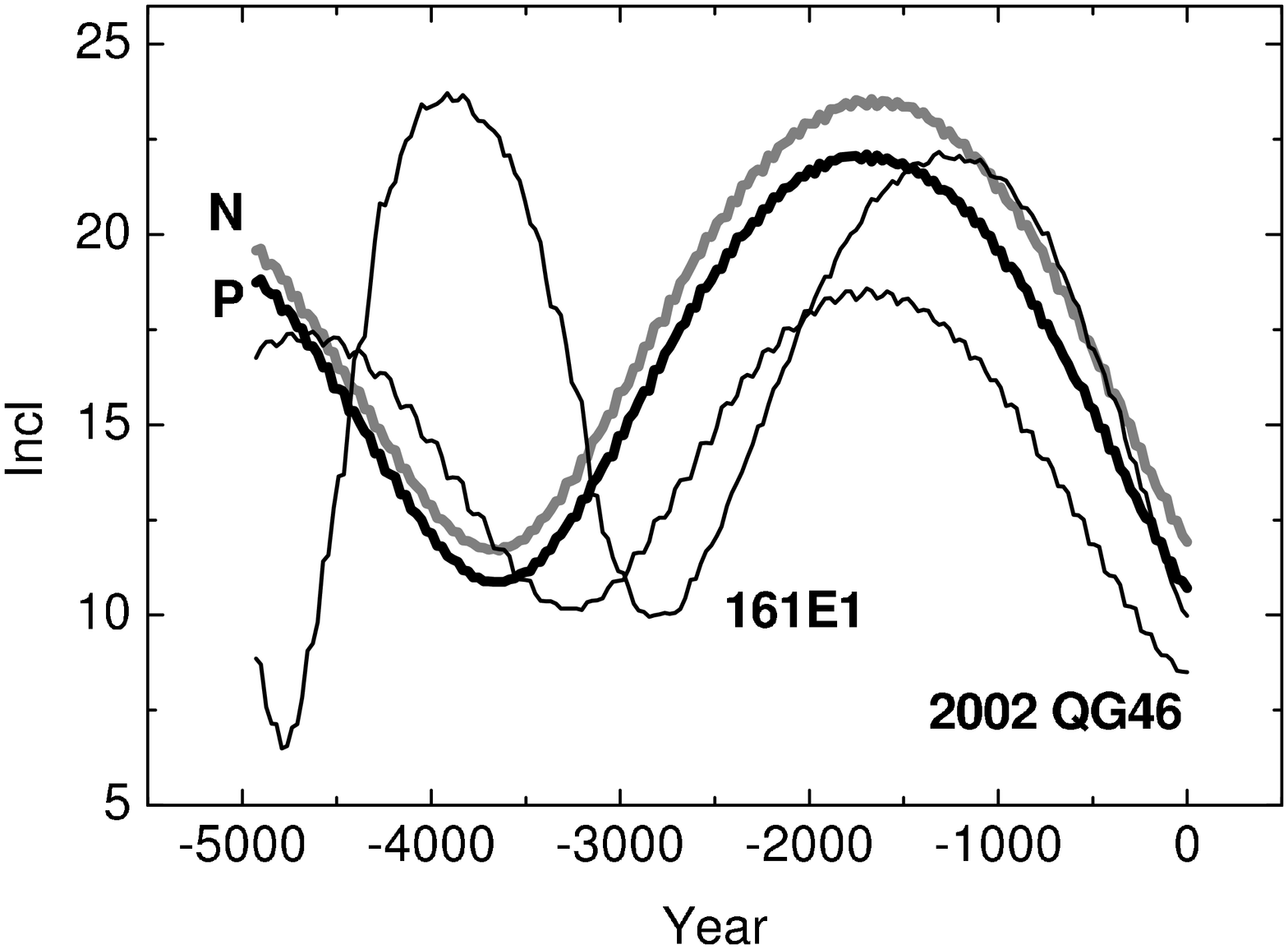}
  \includegraphics[width=0.33\textwidth, angle=-90]{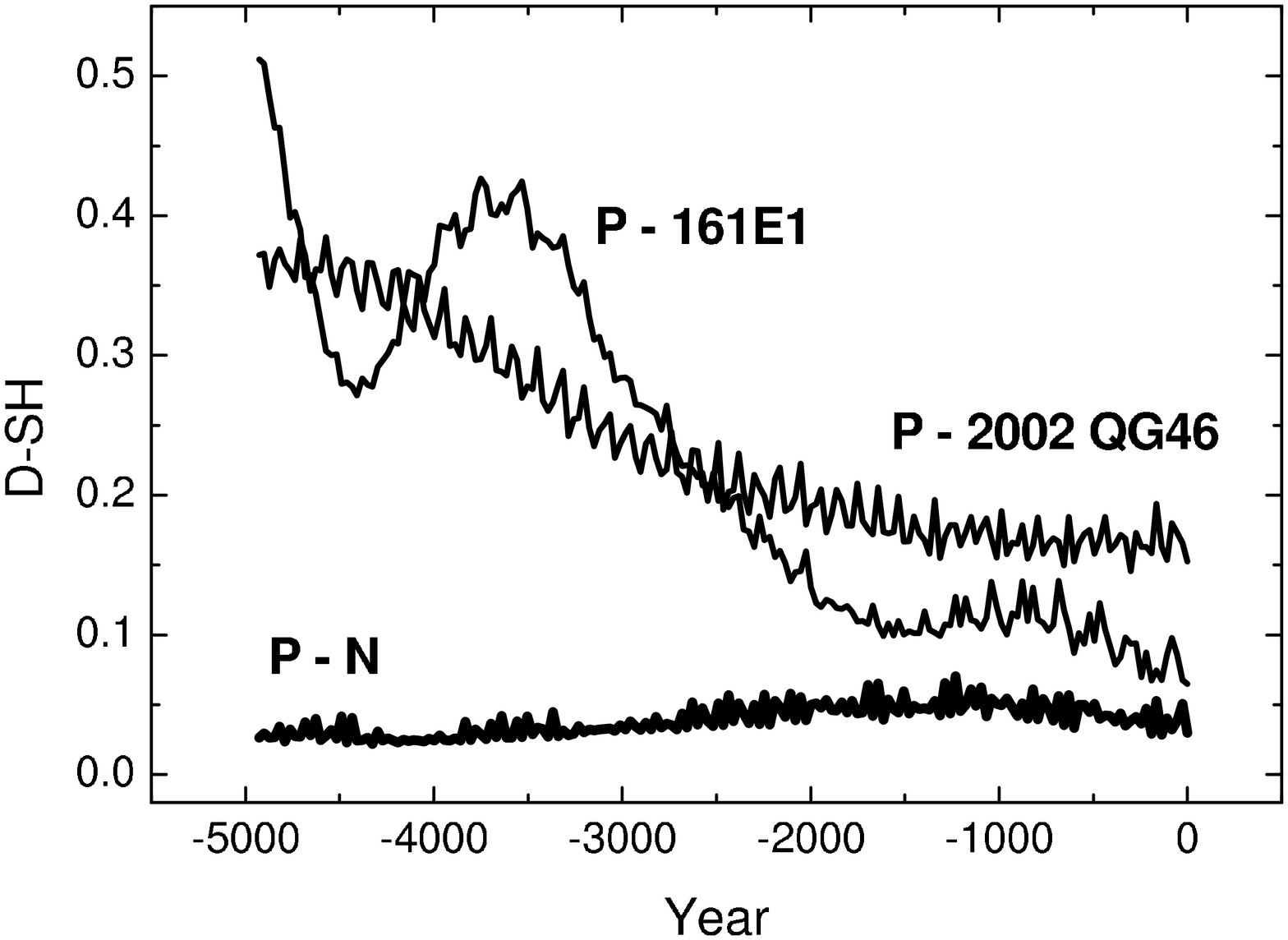}
\vspace{0.1cm} \centering \hspace{0.2cm}
  \includegraphics[width=0.33\textwidth, angle=-90]{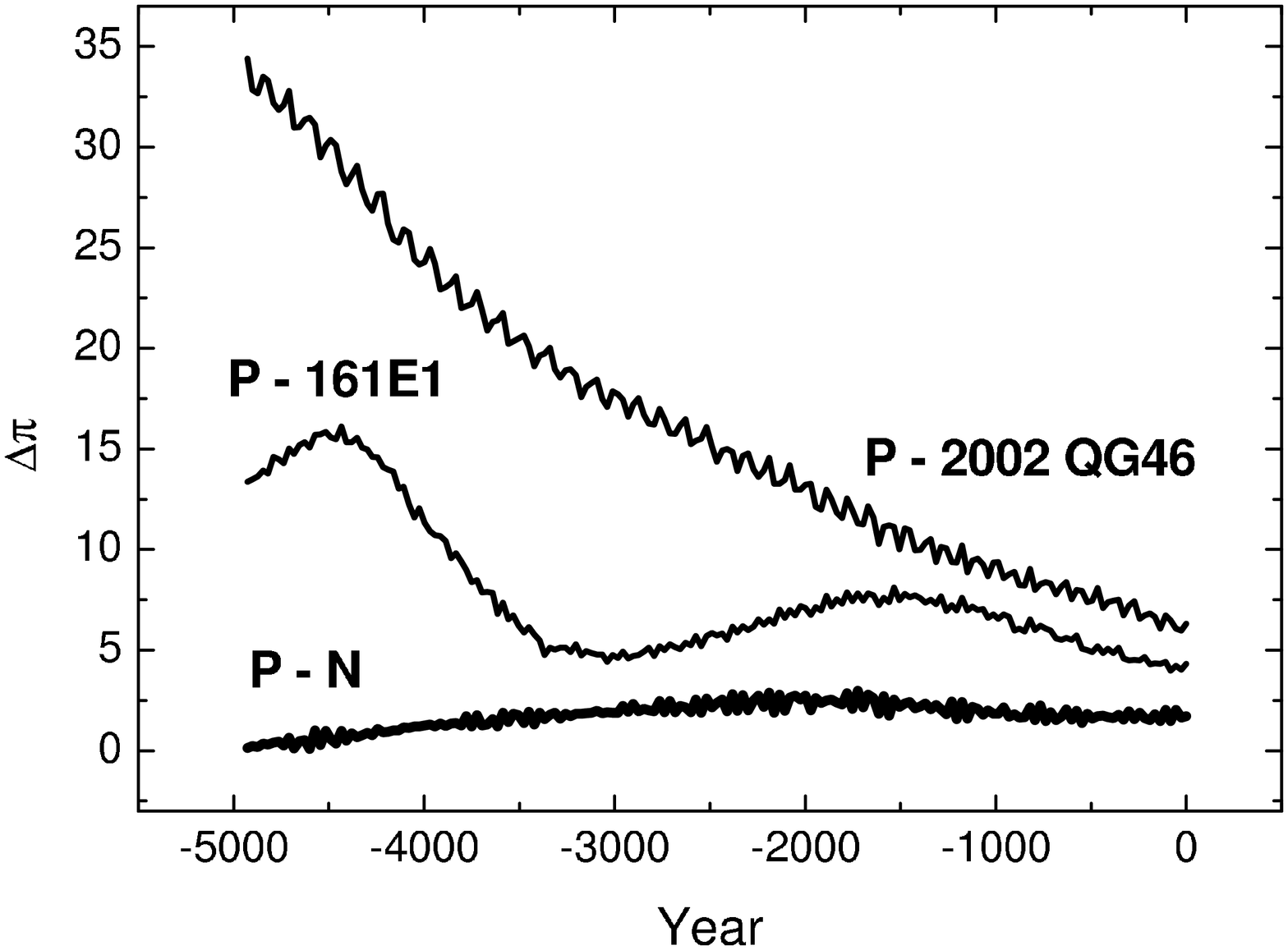}
\caption{The orbital evolution in semimajor axis $\it{a}$, perihelion
distance $\it{q}$,
eccentricity $\it{e}$, inclination $\it{i}$, D-criterion and difference in
longitude of perihelion $\it{\Delta\pi}$ of P\v{r}\'{i}bram (P), Neuschwanstein
(N), meteor 161E1 and asteroid 2002 QG46.} \label{fig:1}
\end{figure*}

Only the orbital evolution of the best associations are presented
in Figure 1. The $D_{SH}$ between P\v{r}\'{i}bram and
Neuschwanstein is within 0.07 and also the difference in the
longitude of perihelion is very small ($\Delta\pi\leq3^\circ$)
during the integration time of 5000 years. This indicates a very
close orbital evolution between the two meteorites. Only one
meteoroid 161E1 and one asteroid 2002 QG46 were found with
reasonably similar evolution to the meteorites in the last 2000
years or so. However, the orbital evolution of asteroid 2002 QG46
is not so close to P\v{r}\'{i}bram. So we prefer only the
meteoroid 161E1 as a possible association.

\section{Clones of P\v{r}\'{i}bram and Neuschwanstein}

Pauls and Gladman (2005) integrated P\v{r}\'{i}bram's orbit for
several hundred thousand years and showed that the substantial
decoherence of the modeled stream occurred in about 50 000
years. However, here we study the orbital evolution of clones covering
the error intervals of P\v{r}\'{i}bram's and Neu\-schwan\-stein's
orbital elements in order to check the stability of their orbital regions.

We have distributed 5 values equidistantly within the error
interval of each parameter (semimajor axis, eccentricity,
inclination, argument of perihelion and mean anomaly). The sixth
parameter, the longitude of node, remained fixed, being of two
orders better precision. Using the combinations of 5 values in 5
orbital parameters, 3125 clones were obtained for each
meteorite.

We have numerically integrated the clones of P\v{r}\'{i}bram and
Neu\-schwan\-stein over the past 5000 years. The orbital evolution
of all clones is more or less similar and stable. The clones of
P\v{r}\'{i}bram are less spread at the end of integration due to
the smaller initial dispersion. The largest dissimilarity in the
orbital evolution is caused by different initial semimajor axes of
clones. A comparison of the orbital evolution of Neuschwanstein
clones that have semimajor axes at the edges of the error
interval ($a=2.38$ AU and $a=2.42$ AU) is presented in Figure 2.
As can be seen, the evolution of both sets of clones is very
similar. Essentially the only difference is that
the period of the variations in perihelion, eccentricity
and inclination for the clones with $a=2.42$ AU is shorter than for
the clones with $a=2.38$ AU. This is caused by the distance of
the orbit from the orbit of Jupiter being smaller, as shown by Wu and
Williams (1992). The descending nodes of almost all clones are
stable and close to the Earth's orbit during the last 3000 years.
The longitude of the ascending node is dispersed by about
$10^\circ$ after 5000 years of evolution. If we suppose that our
clones represent a meteoroid stream, then it would have a similar
dispersion of the orbital elements as that depicted in Figure 2.
The possible stream could be active for at least $\pm$5 days
around the date of the P\v{r}\'{i}bram fall.

\begin{figure*}
\centering
  \includegraphics[width=0.33\textwidth, angle=-90]{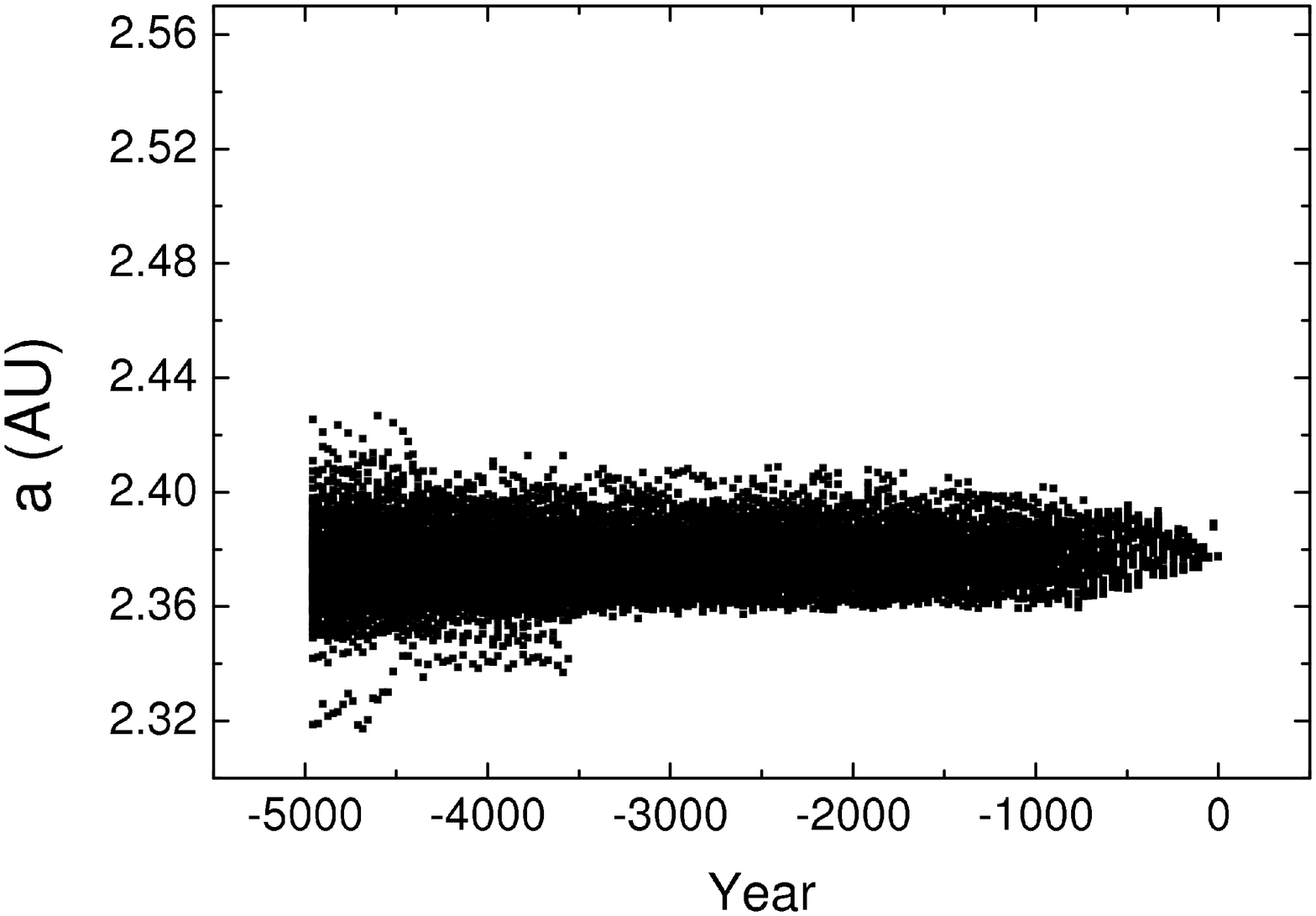}
\vspace{0.3cm} \centering \hspace{0.2cm}
  \includegraphics[width=0.33\textwidth, angle=-90]{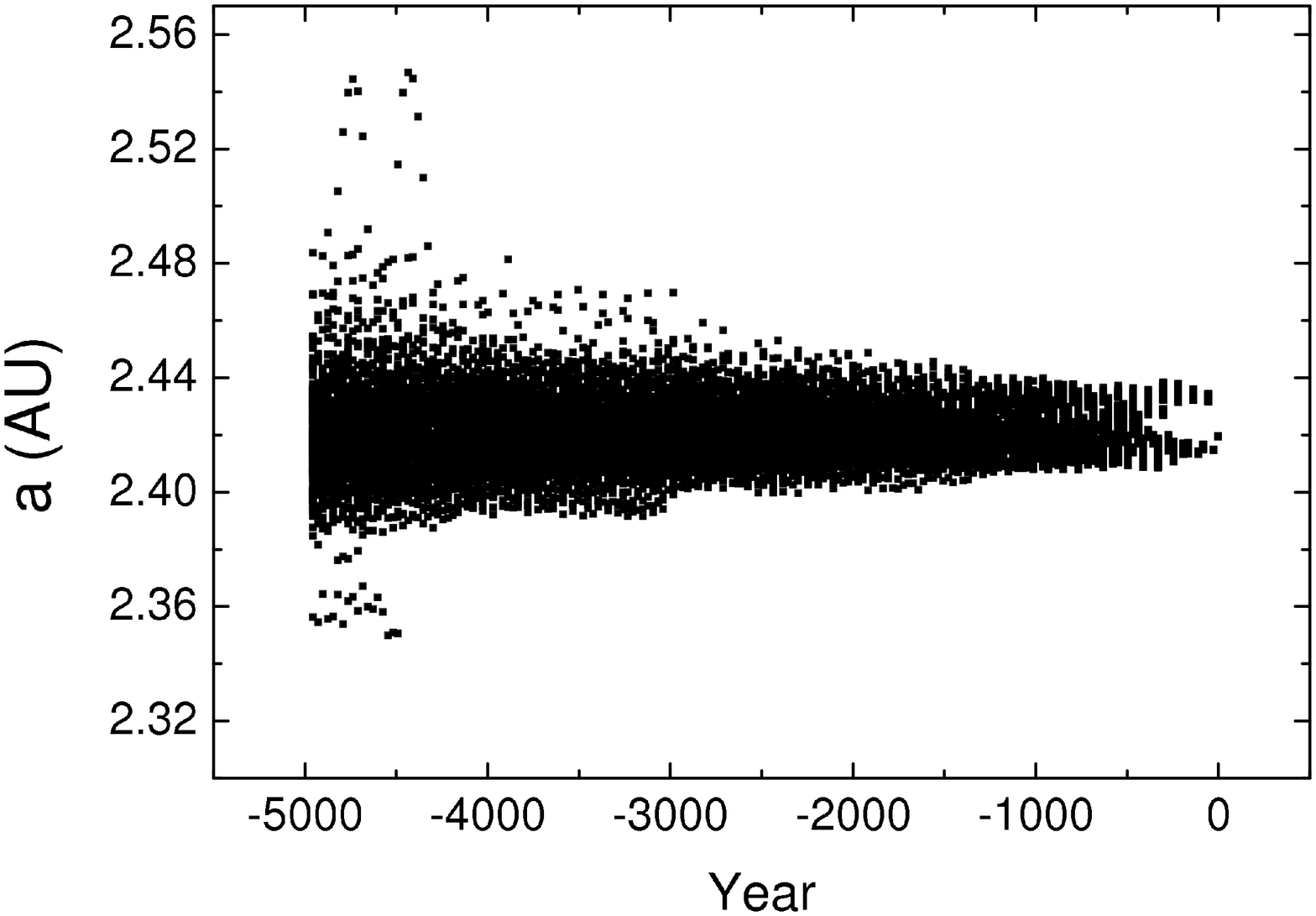}
  \includegraphics[width=0.33\textwidth, angle=-90]{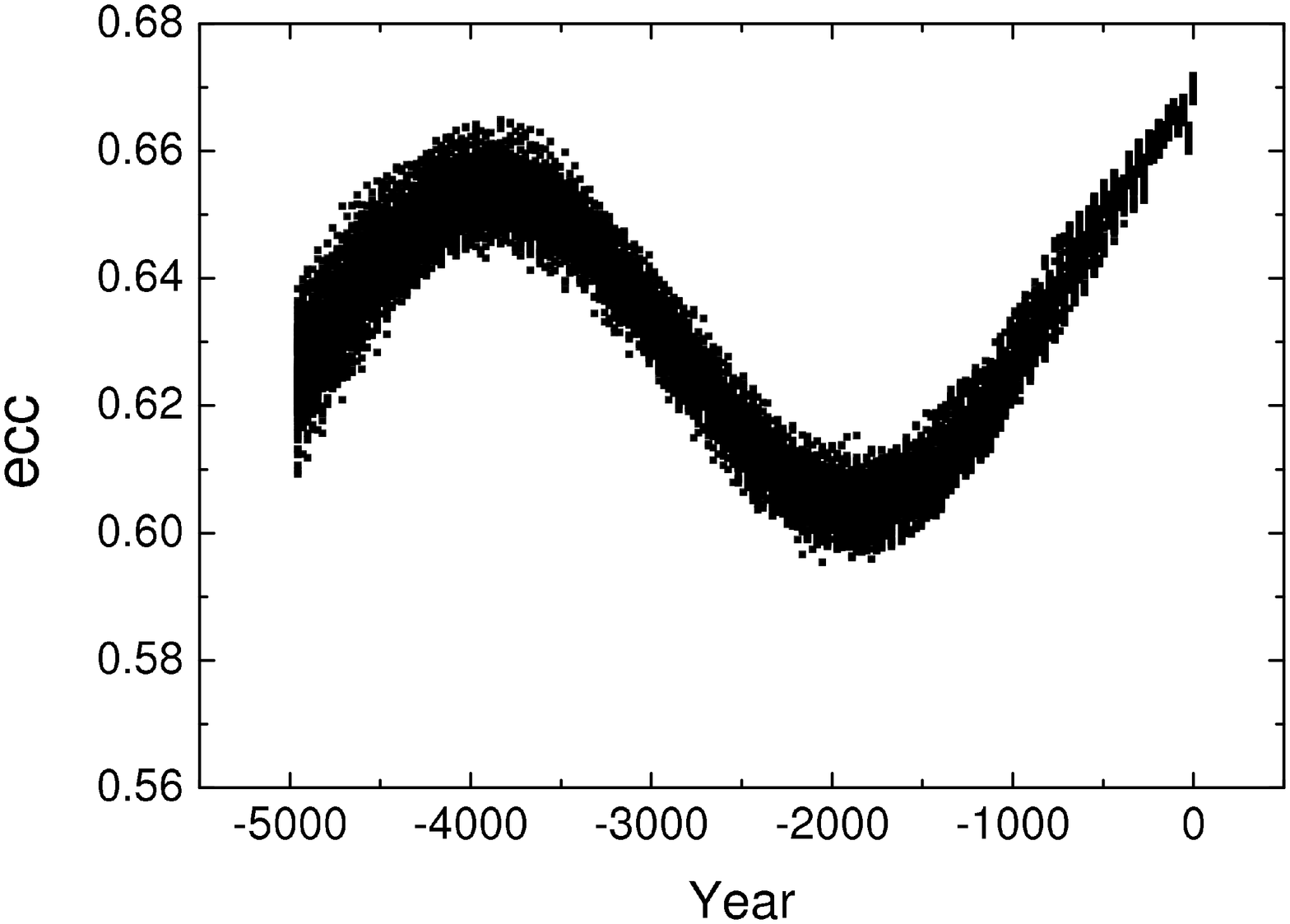}
\vspace{0.3cm} \centering \hspace{0.2cm}
  \includegraphics[width=0.33\textwidth, angle=-90]{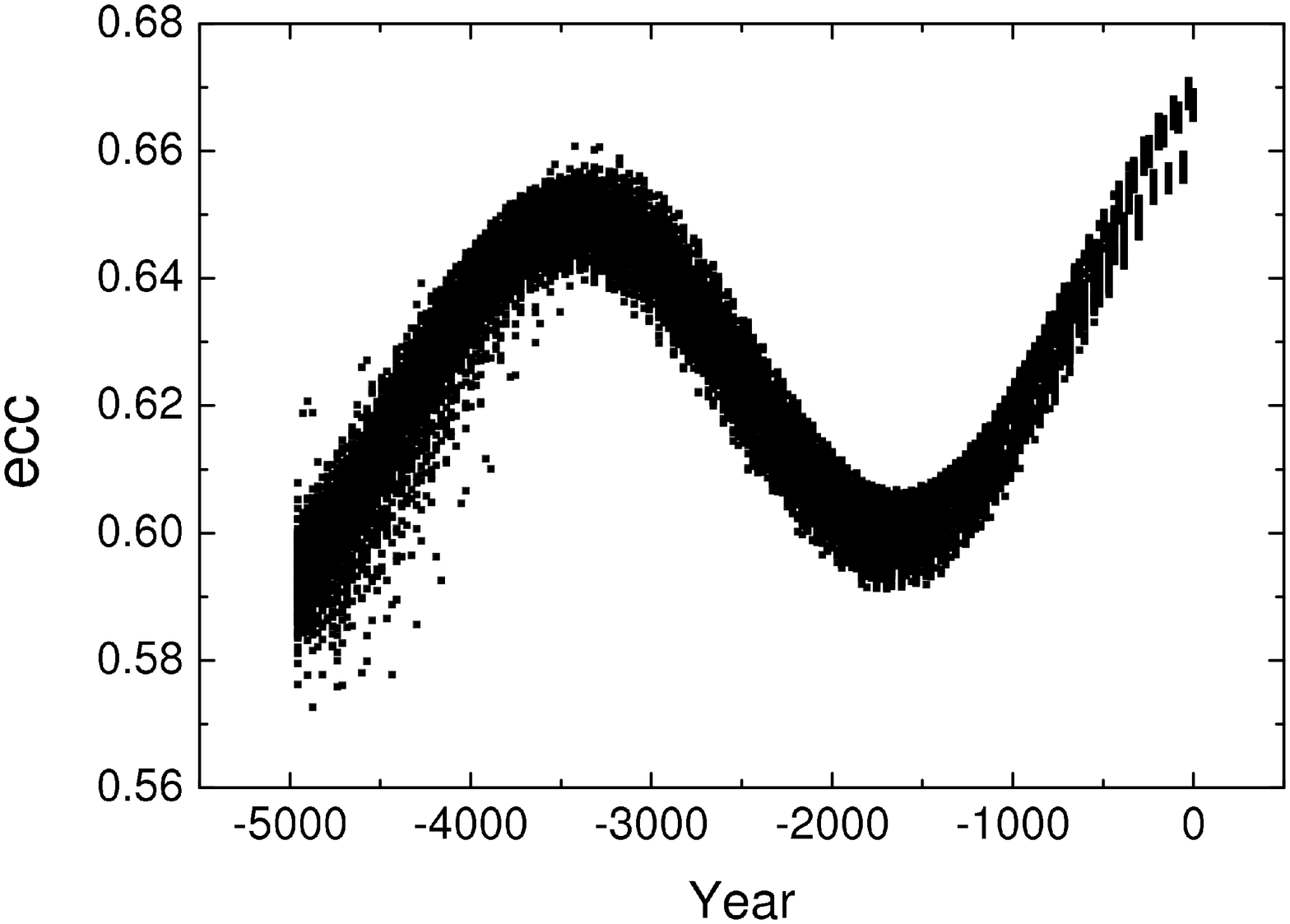}
  \includegraphics[width=0.33\textwidth, angle=-90]{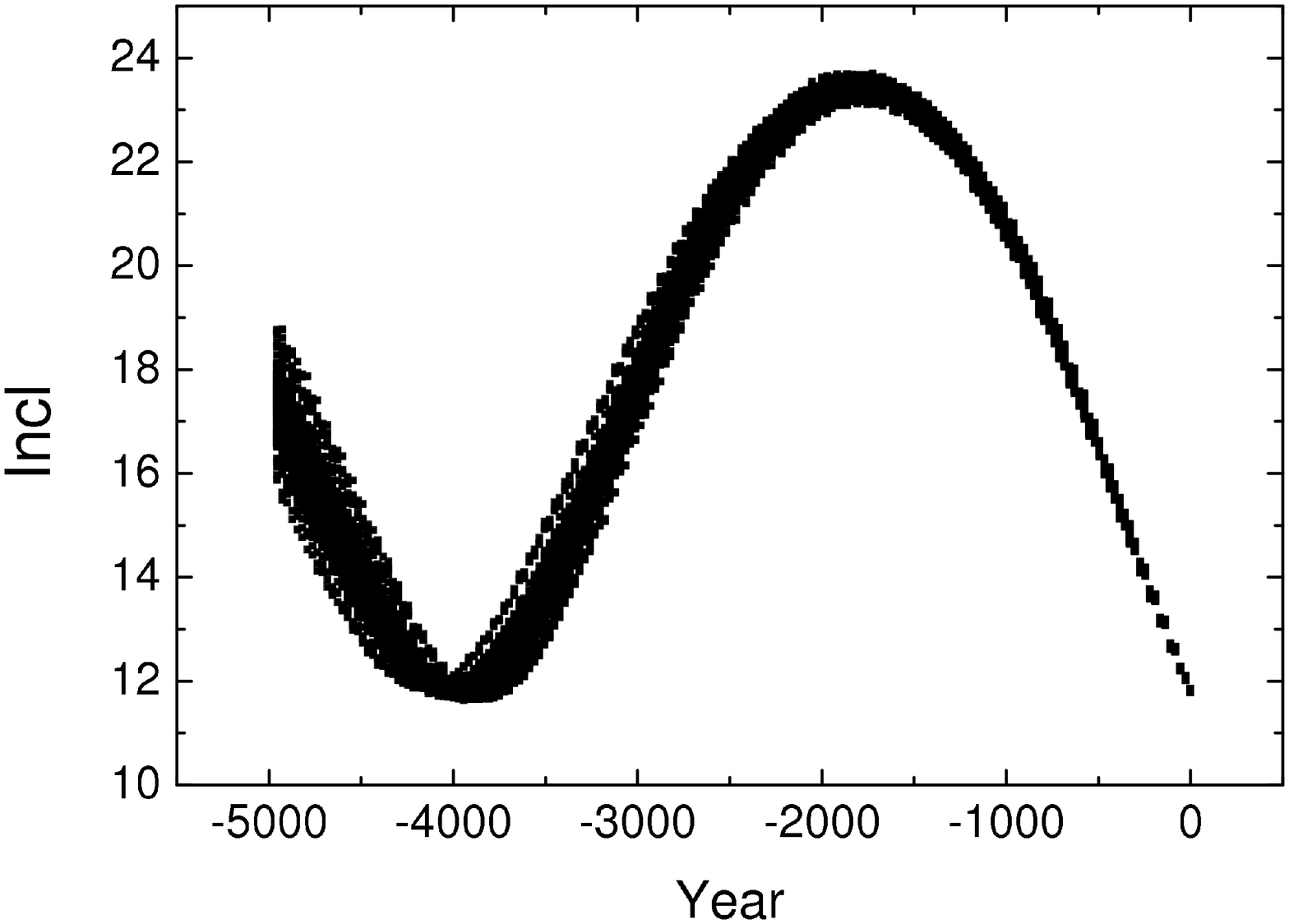}
\vspace{0.3cm} \centering \hspace{0.2cm}
  \includegraphics[width=0.33\textwidth, angle=-90]{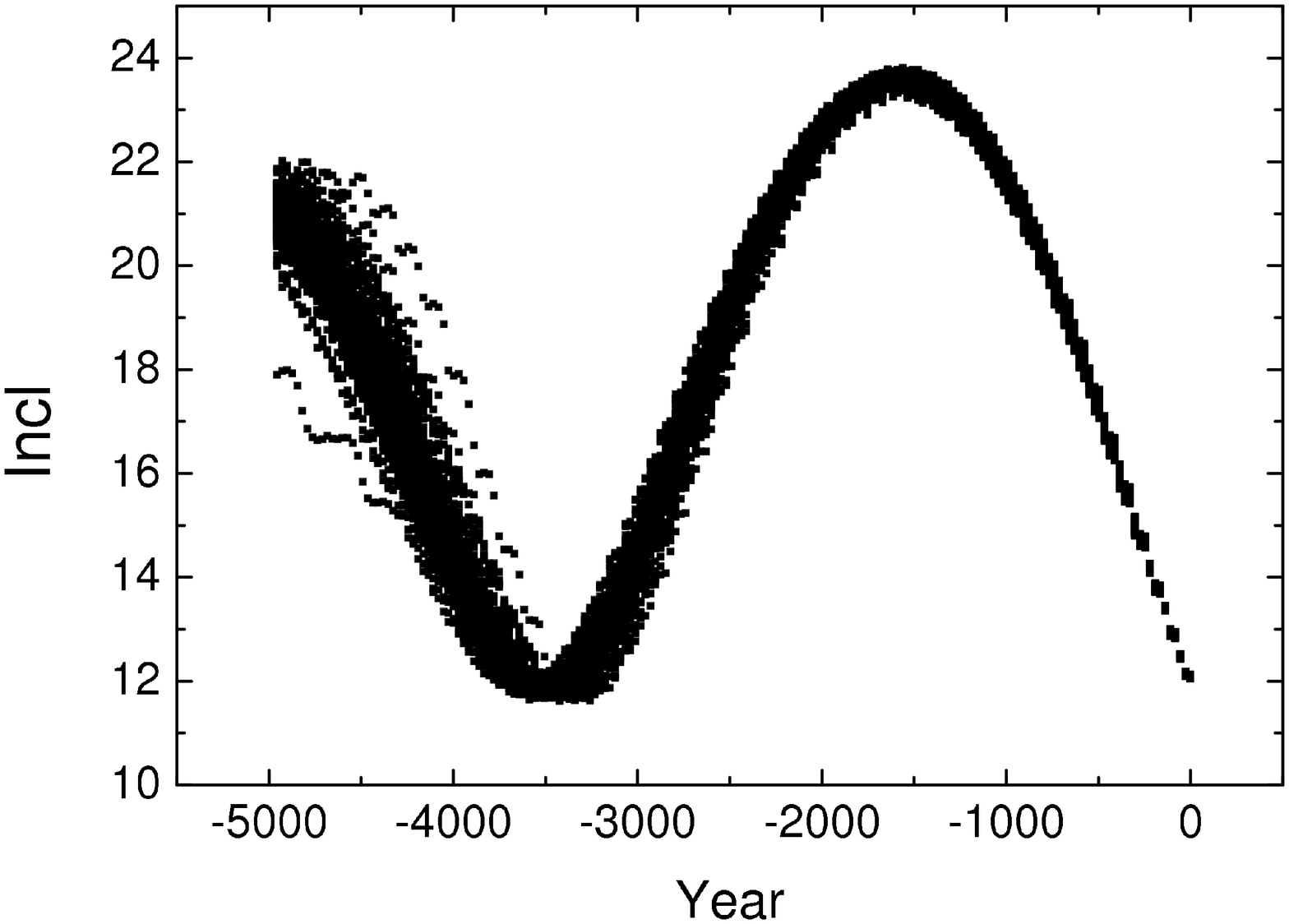}
  \includegraphics[width=0.33\textwidth, angle=-90]{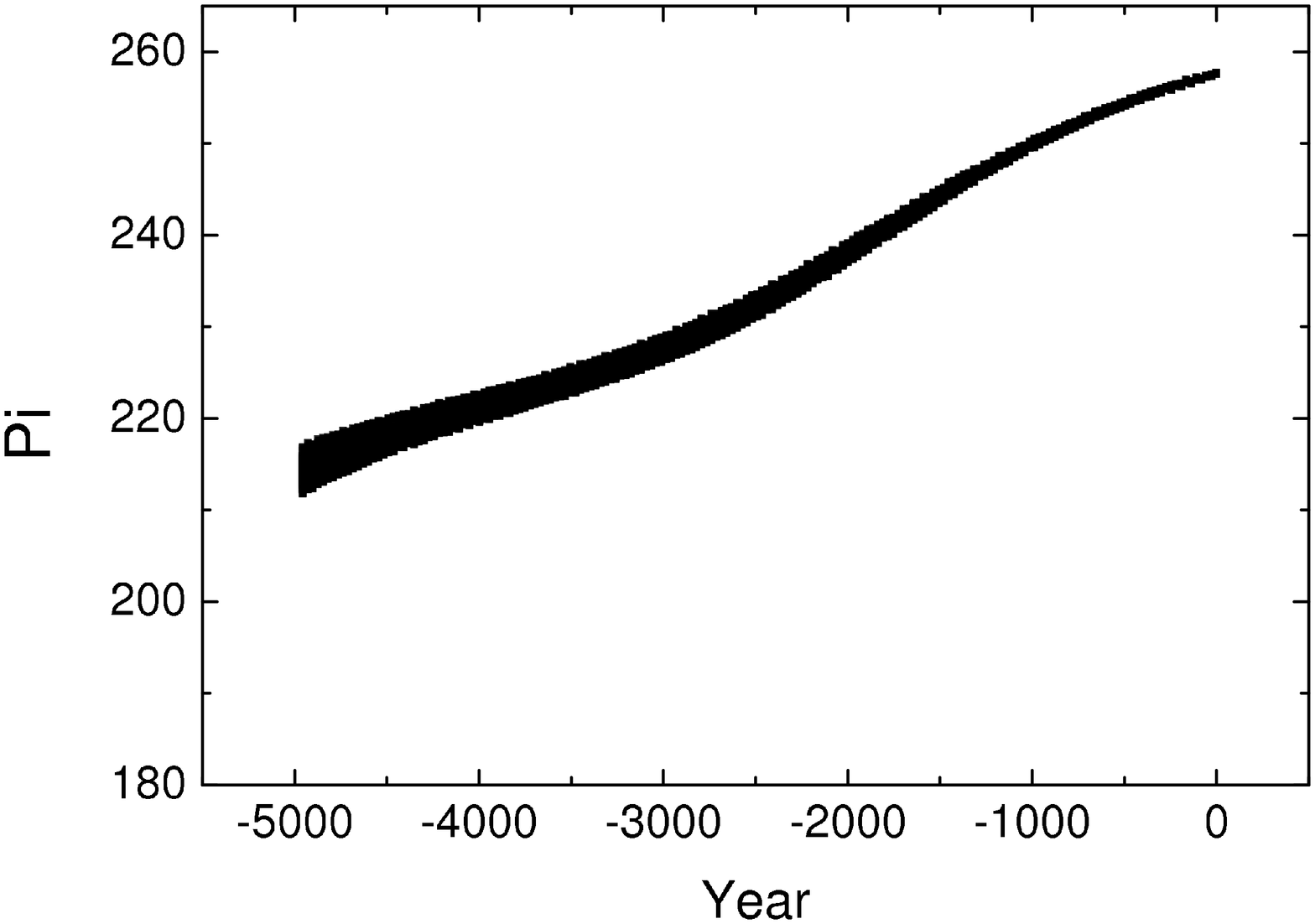}
\vspace{0.1cm} \centering \hspace{0.2cm}
  \includegraphics[width=0.33\textwidth, angle=-90]{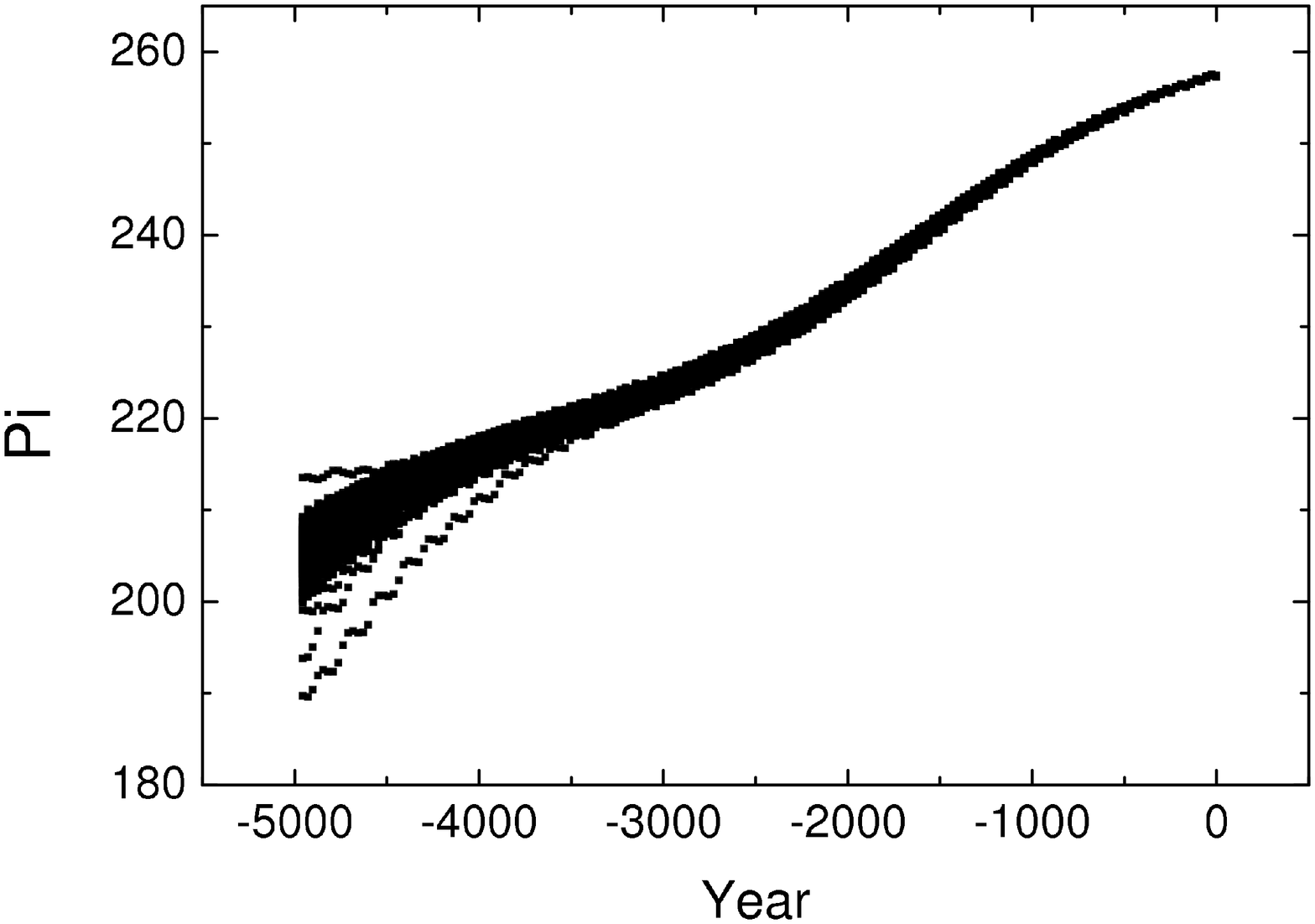}
\caption{The orbital evolution in semimajor axis, eccentricity, inclination and
longitude of perihelion of clones of Neuschwanstein. The left set of graphs
presents the clones for the initial semimajor axis $a=2.38$ AU and the right
set for $a=2.42$ AU.} \label{fig:2}
\end{figure*}

Analysis of the orbital evolution has shown that 75\% of the clones of
P\v{r}\'{i}bram and 84\% of Neuschwanstein experienced close encounters with
the Earth within 0.028 AU in the last 5000 years. This distance is equivalent
to a gravitational perturbation by Jupiter from a distance of 0.5 AU with
respect to the perturbed body. Closer approaches caused a larger spread in the
orbital elements at the end of the integration (Figure 2). Some of the clones
undergo more than one close approach to the Earth. Only a few clones
encountered Mars also.

The results of the orbital integration of the clones of P\v{r}\'{i}bram
and Neuschwanstein show that the orbits are rather stable over
several thousand years. A body with slightly different
orbital elements from P\v{r}\'{i}bram would then also have a similar
evolution. Is it possible that P\v{r}\'{i}bram and Neuschwanstein
have such close orbits by chance?

We are interested in an occurrence of orbits of P\v{r}\'{i}bram
type in a 5 dimensional space of orbital elements. In our previous
paper (Vere\v{s} et al., 2006), we generated and modeled $10^7$
synthetic orbits of 10~m size bodies according to the NEA orbit
distribution of Bottke et al. (2000) and population distribution
of Stuart and Binzel (2004). A probability was found for the
occurrence of each orbital element ($a, e, i, \omega, \Omega$)
within the error boundaries of P\v{r}\'{i}bram and Neuschwanstein.
Then the overall chance of this type of orbit occurring at random
is the product of the probabilities in each element. The resultant
probability is very small, only $2.75\times10^{-11}$.

When we extend this NEA synthetic population to smaller objects, of
the initial radius of the Neu\-schwan\-stein meteoroid 0.3~m (ReVelle
et al., 2004), we obtain a population with a cumulative number
of $2.5\times10^{9}$ (Stuart and Binzel, 2004) or
$1.4\times10^{11}$ (Brown et al., 2002) bodies. Then the expected
occurrence of orbits within the error interval of P\v{r}\'{i}bram and
Neuschwanstein could be from 0.07 to 4 orbits depending on the
real cumulative number in the NEA population.

\section{Conclusions}

If the real number of meteorite producing bodies of size $\sim$ 0.6~m in the
NEA
population is about $10^{11}$, we would expect at least one very close pair in
the P\v{r}\'{i}bram region. This is in good agreement with conclusions of Pauls
and Gladman (2005) that the occurrence of such close orbits is by chance. On
the
other hand, considering a more conservative assessment of $10^{9}$ bodies in
the NEA population, the probability of the existence of the P\v{r}\'{i}bram and
Neuschwanstein pair is very low. Moreover, this probability seems to be even
smaller when we take into account the fact that both bodies entered the Earth's
atmosphere within a time interval of 43 years, as was mentioned by
Spurn\'{y} et al. (2003).

Based on our dynamical investigation described above, we are in
favour of the hypothesis of a common origin of the P\v{r}\'{i}bram
and Neuschwanstein meteorites from a heterogeneous parent
asteroid. The close evolution of the two orbits over several
thousand years is not a proof (e.g. Porub\v{c}an et al. 2004,
Jones et al. 2006, Trigo-Rodr\'{i}guez et al. 2007), but it does
give significant support to suspicions about their common origin.
The parent body of these meteorites could be a rubble pile
asteroid which can possess heterogeneous material gravitationally
aggregated after collisions. In another paper (Vere\v{s} et al.
2007) it has been proposed that relatively recent release of
meteoroids from a parent asteroid by the Earth's tidal force is
possible at substantially larger distances than the Roche limit.
At such distances the differential gravitational influence would
be insufficient to disperse the orbits of released meteoroids from
the parent body. That is why we expect similar orbits of the
parent body and P\v{r}\'{i}bram and Neuschwanstein. We suppose
that the different cosmic-ray ages of the meteorites are affected
by having different cosmic radiation exposure times during which
they were exposed on the surface of the "parent" body.

\begin{acknowledgements}
This work was supported by VEGA - the Slovak Grant Agency for
Science (grant No. 1/3067/06) and by Comenius University grant
UK/401/2007. The authors are grateful to reviewers I.P. Williams and
D. Asher for valuable suggestions.
\end{acknowledgements}

\end{document}